# Wide Field Imaging of van der Waals Ferromagnet Fe$_3$GeTe$_2$ by Spin Defects in Hexagonal Boron Nitride

Mengqi Huang,[1] Jingcheng Zhou,[1] Di Chen,[2,3] Hanyi Lu,[1] Nathan J. McLaughlin,[1] Senlei Li,[1] Mohammed Alghamdi,[4] Dziga Djugba,[1] Jing Shi,[4] Hailong Wang,[5] Chunhui Rita Du[1,5]

[1]Department of Physics, University of California, San Diego, La Jolla, California 92093
[2]Department of Physics, University of Houston, Houston, Texas 77204
[3]Texas Center for Superconductivity, University of Houston, Houston, Texas 77204
[4]Department of Physics and Astronomy, University of California, Riverside, California 92521
[5]Center for Memory and Recording Research, University of California, San Diego, La Jolla, California 92093

Emergent color centers with accessible spins hosted by van der Waals materials have attracted substantial interest in recent years due to their significant potential for implementing transformative quantum sensing technologies. Hexagonal boron nitride (hBN) is naturally relevant in this context due to its remarkable ease of integration into devices consisting of low-dimensional materials. Taking advantage of boron vacancy spin defects in hBN, we report nanoscale quantum imaging of low-dimensional ferromagnetism sustained in Fe$_3$GeTe$_2$/hBN van der Waals heterostructures. Exploiting spin relaxometry methods, we have further observed spatially varying magnetic fluctuations in the exfoliated Fe$_3$GeTe$_2$ flake, whose magnitude reaches a peak value around the Curie temperature. Our results demonstrate the capability of spin defects in hBN of investigating local magnetic properties of layered materials in an accessible and precise way, which can be extended readily to a broad range of miniaturized van der Waals heterostructure systems.

Optically active spin defects in wide band-gap semiconductors promise to enable a broad range of emerging applications in quantum information sciences and technologies.[1–4] To date, nitrogen-vacancy centers in diamond,[1,2] as well as divacancy and silicon-vacancy centers in silicon carbide,[3–5] have been among the most prominent candidates, and have been successfully applied to quantum sensing, computing, and network research, enabling unprecedented field sensitivity, spatial resolution, and state-of-the-art spin-qubit operations.[2,6,7] Many of these advantages derive from the quantum-mechanical nature of these spin defects, which are endowed with excellent quantum coherence, single-spin addressability, and remarkable functionality over a broad temperature range.[1,3–5,8,9]

More recently, the flourishing catalog of van der Waals materials[10] has provided a diverse new playground to enrich this field. There is ongoing and intense activity to explore emergent spin defects and color centers in atomic layers of van der Waals crystals, e. g. transition metal dichalcogenides $MoS_2$,[11,12] $WSe_2$,[12,13] and hexagonal boron nitride (hBN).[12,14–28] In comparison with their conventional counterparts imbedded in three-dimensional solid-state-media, spin defects hosted by two-dimensional (2D) materials exhibit improved versatility for implementing ultrasensitive quantum sensing of proximate objects and remarkable compatibility to device integration.[14,26] For instance, hexagonal boron nitride (hBN), one of the most intensively studied candidates, has been widely employed as an encapsulation layer and gate dielectric material in fabricating functional 2D devices.[29–32] Thus, nanoscale proximity between spin defects in a hBN thin sheet and a layered 2D material can be readily established in van der Waals heterostructures, offering a previously unexploited quantum sensing platform to explore the local physical quantities of interest in an accessible and precise way.

Despite these potential benefits and many pioneering studies, to date, experimental demonstration of quantum microscopy using spin defects in a real van der Waals heterostructure remains a formidable challenge. In this work, we report nanoscale quantum sensing and imaging of exfoliated 2D ferromagnet $Fe_3GeTe_2$ (FGT) flakes[33–41] by boron vacancy $V_B^-$ spin defects in an adjacent hBN capping layer. Exploiting a wide-field magnetometry method,[42–46] we directly image the local magnetic texture of the FGT flake and its characteristic temperature and field dependent magnetization evolution behavior. Taking advantage of spin relaxometry techniques,[47–51] we have observed spatially varying spin fluctuations in the FGT flake, whose magnitude reaches a peak value around the Curie temperature, consistent with the expected ferromagnetic phase transition. We highlight that the presented quantum sensing platform built on spin defects in van der Waals crystals can be extended naturally to a large family of miniaturized 2D heterostructure systems,[10,52,53] bringing new opportunities for investigating the local spin, charge, and thermal properties of emergent quantum materials and devices.

Before discussing the details of our experimental results, we first review our measurement platform and device structure, as illustrated in Fig. 1a. An exfoliated FGT flake is mechanically transferred onto a patterned Au microwave transmission line, followed by encapsulation with an hBN layer. The device preparation process was performed in a glove box with argon environment to minimize environmental effects (see Methods for details). An optical microscope image shown in Fig. 1b provides an overview of a prepared device, where the thickness of the FGT and hBN flakes were characterized to be 86 and 88 nm, respectively (see Supplementary Information Note 1 for details). Boron vacancy $V_B^-$ spin defects in the hBN flake were created by Helium ion implantation. Figure 1c illustrates the structure of $V_B^-$ in a hexagonal crystalline structure with alternating boron (red) and nitrogen (green) atoms, where three nitrogen atoms are adjacent to each boron atom vacancy $V_B^-$. The negatively charged $V_B^-$ spin defect has an $S = 1$ electron spin and serves as a three-level quantum system, as shown in Fig. 1d. In the present study, microwave currents flowing in the Au microwave transmission line were used to control the quantum spin state of $V_B^-$ spin defects, which can then be optically accessed via spin-dependent photoluminescence. The Au underlayer also enhances the photoluminescence and optical contrast of $V_B^-$ spin defects,[16] aiding the quantum microscopy measurements discussed below. The distance between the $V_B^-$ defect centers and the top surface of the FGT sample is estimated to be ~50 nm based on Stopping and Range of Ions in Matter (SRIM) simulations. Exfoliated FGT flakes show spontaneous perpendicular magnetization due to reduced crystal symmetry of the layered structure.[33,35,36] We have prepared separate devices to systematically and reproducibly characterize the magneto-transport properties of exfoliated FGT flakes, whose Curie temperature is measured to be ~200 K, in qualitative agreement with previous studies (see Supplementary Information Note 2 for details).[33,35,40]

We now utilize wide-field microscopy to demonstrate optically detected magnetic resonance (ODMR) and spin relaxation of $V_B^-$ defect centers (see Methods for details). The top panel of Fig. 1e shows the optical and microwave measurement sequence. We utilize 1-μs-long green laser pulses for spin initialization and readout, and ~100-ns-long microwave π pulses[54] to induce spin transitions of $V_B^-$ spin defects. We sweep the frequency $f$ of the microwave π pulses and measure the fluorescence across the field of view projected on a CMOS camera. When $f$ matches the electron spin resonance (ESR) frequencies, $V_B^-$ spin defects are excited to the $m_s = \pm 1$ states, which are more likely to relax through a non-radiative pathway back to the $m_s = 0$ ground state and emit reduced photoluminescence (PL). The bottom panel of Fig. 1e shows a series of ODMR spectra of $V_B^-$ defect centers measured at room temperature with different external magnetic fields $B_{ext}$ applied along the out-of-plane direction. For $B_{ext} = 0$, the energy level of the $m_s = \pm 1$ states of $V_B^-$ spin defects exhibit a small separation of ~100 MHz due to the off-axial zero field splitting effect[14,26,55] and the average ESR frequency equals ~3.47 GHz at room temperature.[14,26,28,55] For $B_{ext} > 0$, the Zeeman coupling separates the $m_s = -1$ and $m_s = +1$ spin states by an energy gap of magnitude $2\tilde{\gamma}B_{ext}$ (see Supplementary Information Note 3 for details), where $\tilde{\gamma}$ denotes the gyromagnetic ratio of $V_B^-$ defect centers. To characterize the quantum coherence of $V_B^-$ spin defects, we perform spin relaxometry measurements with a measurement protocol shown in the top panel of Fig. 1f. A microsecond scale green laser pulse is first applied to initialize the $V_B^-$ spin defects to the $m_s = 0$ state. During the time delay, fluctuating magnetic fields at the ESR frequencies will accelerate spin relaxation of $V_B^-$ spin defects. After a delay time $t$, we measure the occupation probabilities of the $V_B^-$ spin defects at the $m_s = 0$ and $\pm 1$ states by applying a microwave π pulse on the corresponding ESR frequencies and measuring the spin-dependent PL by a green-laser readout pulse. The bottom panel of Fig. 1f shows the integrated PL intensity of the $m_s = 0$ and $\pm 1$ states measured as a function of the delay time $t$. By fitting the data with a three-level model,[51,56] the spin relaxation rate $\Gamma_0$ of $V_B^-$ spin defects is obtained to be 39 KHz (38 KHz) for $m_s = 0 \rightarrow -1 (+1)$ transition at 295 K, consistent with previous studies.[16,21,55]

After demonstrating the ODMR and spin relaxometry measurement capabilities, next, we use $V_B^-$ spin defects in hBN to directly image magnetic textures of an exfoliated FGT flake. Wide-field magnetometry exploits the Zeeman splitting effect of the ensembles of $V_B^-$ defect centers to measure the local magnetic stray fields generated from the proximate FGT flake, as illustrated in Fig. 2a. It is worth noting that the $V_B^-$ spins are naturally orientated along the out-of-plane direction,[14] serving as an ideal local sensor to investigate the magnetic dynamics and phase transition of FGT with spontaneous perpendicular anisotropy. The magnitude of the local static magnetic field $B_{tot}$ can be extracted as follows: $B_{tot} = \pi \Delta f_{ESR}/\tilde{\gamma}$, where $\Delta f_{ESR}$ characterizes the Zeeman splitting of the $V_B^-$ spin defects. Subtracting the contribution from the external magnetic field $B_{ext}$, the magnetic stray field $B_F$ generated from the FGT flake can be quantitatively measured. By performing spatially dependent ODMR measurements over the $V_B^-$ spin ensembles, we are able to obtain a 2D stray field map as shown in Fig. 2b, which is measured at a temperature $T = 6$ K and an external magnetic field $B_{ext} = 142$ G. Through well-established reverse-propagation protocols[46,57,58] (see Supplementary Information Note 3 for details), the corresponding magnetization $4\pi M$ map of the FGT flake can be reconstructed, as shown in Fig. 2c. The spatially averaged magnetization of the FGT flake is 1.3 kG at 6 K, in qualitative agreement with the bulk value.[33] The variation of the local magnetization could result from inhomogeneities, magnetic domains, or localized defects.[39,59]

We now present systematic wide-field magnetometry results to directly image the magnetic phase transition of the FGT flake across the Curie temperature. Figures 3a-3g show the reconstructed magnetization maps of the FGT flake measured with temperatures varying from 6 to 225 K and an external perpendicular magnetic field $B_{ext}$ of 142 G. In the low temperature regime ($T < 100$ K), the exfoliated FGT flake exhibits robust magnetization, as shown in Figs. 3a and 3b, indicating a long-range ferromagnetic order sustained by the intrinsic magnetocrystalline anisotropy of FGT. The measured FGT magnetization decreases with increasing temperature due to enhanced thermal fluctuations (Fig. 3c). When approaching the Curie temperature, the energy of thermal fluctuations becomes comparable to the exchange energy of FGT, resulting in significant suppression of the FGT magnetization (Fig. 3d). Further increasing of the temperature leads to shrinking of the magnetic domain and decreasing magnetization of the FGT flake (Figs. 3e and 3f). Above the Curie temperature ($T = 225$ K), measured magnetization disappears over the entire FGT flake area (Fig. 3g). Figure 3h summarizes the temperature-dependent evolution of the spatially averaged magnetization of the FGT flake. The magnetic moment of the FGT flake exhibits a gradual decay in the low temperature regime, followed by a dramatic drop during the ferromagnetic phase transition, in agreement with the magneto-transport characterization results (see Supplementary Information Note 2 for details). To further highlight the evolution of magnetic domains in the FGT flake, we present wide-field magnetometry results under different external magnetic fields. Figures 4a-4g show a series of magnetization maps of the FGT flake measured with $B_{ext}$ varied from 142 G to 698 G at a fixed temperature of 178 K. Qualitatively, the FGT magnetization increases with increasing $B_{ext}$ and reaches a saturation value when $B_{ext} \geq 600$ G. This is accompanied by the propagation of magnetic domain walls and expansion of magnetic domain taking place at the nanoscale as shown in the presented images. Figure 4h shows field dependence of the spatially averaged FGT magnetization, consistent with the variation of the anomalous Hall resistance measured in the same magnetic field regime (see Supplementary Information Note 2 for details).

In addition to sensing static magnetic stray fields, the excellent quantum coherence of $V_B^-$ spin defects in hBN also provides the opportunity for probing fluctuating magnetic fields that are challenging to access by conventional magnetometry methods.[47,48,50,51,55] Lastly, we utilize the spin relaxometry method as demonstrated above to probe the temperature dependence of the magnetic fluctuations in the FGT flake, revealing the intriguing physics underlying the longitudinal magnetic susceptibility and diffusive spin transport properties.[60,61] Figures 5a-5g show a series of $V_B^-$ spin relaxation rate maps measured with temperatures between 165 K and 223 K. Note that the background of the intrinsic relaxation rate of $V_B^-$ has been subtracted to highlight the contribution $\Gamma_M$ from the fluctuating magnetic fields generated by FGT (see Supplementary Information Note 4 for details). Due to the strong perpendicular magnetic anisotropy, the minimum magnon energy of FGT is larger than the ESR frequencies of $V_B^-$ spin defects in our experimentally accessible magnetic field range, hence, the measured spin relaxation is driven by the longitudinal spin fluctuations of FGT, which is further related to the static longitudinal magnetic susceptibility $\chi_0$ and the diffusive spin transport constant $D$.[46,60,61] When temperature is away from the quantum critical point, magnetic fluctuations in FGT are largely suppressed due to its vanishingly small magnetic susceptibility, leading to negligible spin relaxation rate $\Gamma_M$ of $V_B^-$ defects (Figs. 5a and 5b). In contrast, we observed significantly enhanced spin relaxation rate during the magnetic phase transition of FGT (Figs. 5c-5e), which we attribute to the increase of the magnetic susceptibility of FGT around the Curie temperature.[62] When $T$ is above 200 K, spin fluctuations remain active in FGT due to the finite spin-spin correlation in the paramagnetic state.[63] The observed spatially varying magnetic fluctuations over the exfoliated FGT flake could be induced by inhomogeneities in magnetic susceptibility, spin diffusion constant, and exchange coupling strength. Figure 5h summarizes the temperature dependence of the spatially averaged spin relaxation rate $\Gamma_M$ with a peak value of 36 kHz around the Curie temperature, consistent with the ferromagnetic phase transition of FGT.[62] Invoking a theoretical model developed in Ref. 60, the longitudinal magnetic susceptibility $\chi_0$ and spin diffusion constant $D$ of the exfoliated FGT flake is extracted to be $(1.5 \pm 0.2) \times 10^{-2}$ emu·cm$^{-3}$·Oe$^{-1}$ and $(1.7 \pm 0.3) \times 10^{-5}$ m$^2$/s at 189 K (see Supplementary Information Note 4 for details).

In summary, we have demonstrated $V_B^-$ spin defects in hBN as a local probe to image magnetic phase transitions and spin fluctuations in the archetypical van der Waals ferromagnet FGT at the nanoscale. The spatially resolved wide-field magnetometry results reveal the characteristic evolution behavior of magnetic domains during the phase transition of FGT. By using $V_B^-$ spin relaxometry techniques, we are also able to access the spin fluctuations in the FGT flake, whose magnitude reaches a maximum value around the Curie temperature. Our results illustrate the appreciable capability of $V_B^-$ spin defects hosted by hBN of investigating local magnetic properties of layered materials in van der Waals heterostructure formats. While the current study is conducted using wide-field magnetometry methods with a spatial resolution set by the optical diffraction limit, we anticipate that the spatial sensitivity of $V_B^-$ spin defects could ultimately reach the tens of nanometers regime by utilizing single-spin defects and developing scanning microscopy measurement schemes,[2,44,50,57,64,65] providing new opportunities to uncover detailed microscopic features in a broad range of 2D material systems.

**Acknowledgements**. Authors would like to thank Eric E. Fullerton for providing the Physical Property Measurement System (PPMS) for electrical transport measurements and Yuxuan Xiao for assistance in sample preparation and magnetometry characterization. Authors thank Shu Zhang for insightful discussions. J. Z., H. L., N. J. M., H. W. and C. R. D. were supported by the Air Force Office of Scientific Research under award FA9550-20-1-0319 and its Young Investigator Program under award FA9550-21-1-0125. M. H., S. L., D. D. and C. R. D. acknowledged the support from U. S. National Science Foundation (NSF) under award ECCS-2029558 and DMR-2046227. M. A. and J. S. were supported by NSF under award ECCS-2051450.

# References

1. Doherty, M. W. et al. The nitrogen-vacancy colour centre in diamond. *Phys. Rep.* **528**, 1–45 (2013).
2. Degen, C. L., Reinhard, F. & Cappellaro, P. Quantum sensing. *Rev. Mod. Phys.* **89**, 035002 (2017).
3. Koehl, W. F., Buckley, B. B., Heremans, F. J., Calusine, G. & Awschalom, D. D. Room temperature coherent control of defect spin qubits in silicon carbide. *Nature* **479**, 84–87 (2011).
4. Nagy, R. et al. High-fidelity spin and optical control of single silicon-vacancy centres in silicon carbide. *Nat. Commun.* **10**, 1954 (2019).
5. Widmann, M. et al. Coherent control of single spins in silicon carbide at room temperature. *Nat. Mater.* **14**, 164–168 (2015).
6. Childress, L. & Hanson, R. Diamond NV centers for quantum computing and quantum networks. *MRS Bull.* **38**, 134–138 (2013).
7. Yao, N. Y. et al. Scalable architecture for a room temperature solid-state quantum information processor. *Nat. Commun.* **3**, 800 (2012).
8. Rose, B. C. et al. Observation of an environmentally insensitive solid-state spin defect in diamond. *Science* **361**, 60–63 (2018).
9. Christle, D. J. et al. Isolated spin qubits in SiC with a high-fidelity infrared spin-to-photon interface. *Phys. Rev. X* **7**, 021046 (2017).
10. Geim, A. K. & Grigorieva, I. V. Van der Waals heterostructures. *Nature* **499**, 419–425 (2013).
11. Ye, M., Seo, H. & Galli, G. Spin coherence in two-dimensional materials. *Npj Comput. Mater.* **5**, 44 (2019).
12. Aharonovich, I., Englund, D. & Toth, M. Solid-state single-photon emitters. *Nat. Photonics* **10**, 631–641 (2016).
13. Chakraborty, C., Kinnischtzke, L., Goodfellow, K. M., Beams, R. & Vamivakas, A. N. Voltage-controlled quantum light from an atomically thin semiconductor. *Nat. Nanotechnol.* **10**, 507–511 (2015).
14. Gottscholl, A. et al. Initialization and read-out of intrinsic spin defects in a van der Waals crystal at room temperature. *Nat. Mater.* **19**, 540–545 (2020).
15. Healey, A. J. et al. Quantum microscopy with van der Waals heterostructures. Preprint at https://arxiv.org/abs/2112.03488 (2021).
16. Gao, X. et al. High-contrast plasmonic-enhanced shallow spin defects in hexagonal boron nitride for quantum sensing. *Nano Lett.* **21**, 7708–7714 (2021).
17. Mendelson, N. et al. Coupling spin defects in a layered material to nanoscale plasmonic cavities. *Adv. Mater.* 2106046 (2021).
18. Yu, P. et al. Excited-state spectroscopy of spin defects in hexagonal boron nitride. Preprint at https://arxiv.org/abs/2112.02912 (2021).
19. Mathur, N. et al. Excited-state spin-resonance spectroscopy of $V_B^-$ defect centers in hexagonal boron nitride. Preprint at https://arxiv.org/abs/2111.10855 (2021).
20. Baber, S. et al. Excited state spectroscopy of boron vacancy defects in hexagonal boron nitride using time-resolved optically detected magnetic resonance. Preprint at https://arxiv.org/abs/2111.11770 (2021).
21. Haykal, A. et al. Decoherence of $V_B^-$ spin defects in monoisotopic hexagonal boron nitride. Preprint at https://arxiv.org/abs/2105.12317 (2021).
22. Murzakhanov, F. F. et al. Electron-nuclear coherent coupling and nuclear spin readout through optically polarized $V_B^-$ spin states in hBN. Preprint at https://arxiv.org/abs/2112.10628 (2021).

23. Tran, T. T., Bray, K., Ford, M. J., Toth, M. & Aharonovich, I. Quantum emission from hexagonal boron nitride monolayers. *Nat. Nanotechnol.* **11**, 37–41 (2016).
24. Grosso, G. et al. Tunable and high-purity room temperature single-photon emission from atomic defects in hexagonal boron nitride. *Nat. Commun.* **8**, 705 (2017).
25. Su, C. et al. Tuning color centers at a twisted interface. Preprint at https://arxiv.org/abs/2108.04747. (2021)
26. Gottscholl, A. et al. Spin defects in hBN as promising temperature, pressure and magnetic field quantum sensors. *Nat. Commun.* **12**, 4480 (2021).
27. Xu, X. et al. Creating quantum emitters in hexagonal boron nitride deterministically on chip-compatible substrates. *Nano Lett.* **21**, 8182–8189 (2021).
28. Liu, W. et al. Temperature-dependent energy-level shifts of spin defects in hexagonal boron nitride. *ACS Photonics* **8**, 1889–1895 (2021).
29. Yankowitz, M. et al. Tuning superconductivity in twisted bilayer graphene. *Science* **363**, 1059–1064 (2019).
30. Jiang, S., Shan, J. & Mak, K. F. Electric-field switching of two-dimensional van der Waals magnets. *Nat. Mater.* **17**, 406–410 (2018).
31. Huang, B. et al. Electrical control of 2D magnetism in bilayer $CrI_3$. *Nat. Nanotechnol.* **13**, 544–548 (2018).
32. Sanchez-Yamagishi, J. D. et al. Helical edge states and fractional quantum Hall effect in a graphene electron–hole bilayer. *Nat. Nanotechnol.* **12**, 118–122 (2017).
33. Deng, Y. et al. Gate-tunable room-temperature ferromagnetism in two-dimensional $Fe_3GeTe_2$. *Nature* **563**, 94–99 (2018).
34. Tu, Z. et al. Ambient effect on the Curie temperatures and magnetic domains in metallic two-dimensional magnets. *Npj 2D Mater. Appl.* **5**, 62 (2021).
35. Fei, Z. et al. Two-dimensional itinerant ferromagnetism in atomically thin $Fe_3GeTe_2$. *Nat. Mater.* **17**, 778–782 (2018).
36. Wang, X. et al. Current-driven magnetization switching in a van der Waals ferromagnet $Fe_3GeTe_2$. *Sci. Adv.* **5**, eaaw8904 (2019).
37. Yang, M. et al. Creation of skyrmions in van der Waals ferromagnet $Fe_3GeTe_2$ on $(Co/Pd)_n$ superlattice. *Sci. Adv.* **6**, eabb5157 (2020).
38. Alghamdi, M. et al. Highly efficient spin–orbit torque and switching of layered ferromagnet $Fe_3GeTe_2$. *Nano Lett.* **19**, 4400–4405 (2019).
39. Zhang, K. et al. Gigantic current control of coercive field and magnetic memory based on nanometer-thin ferromagnetic van der Waals $Fe_3GeTe_2$. *Adv. Mater.* **33**, 2004110 (2021).
40. Xu, J., Phelan, W. A. & Chien, C.-L. Large anomalous Nernst effect in a van der Waals ferromagnet $Fe_3GeTe_2$. *Nano Lett.* **19**, 8250–8254 (2019).
41. Wu, Y. et al. Néel-type skyrmion in $WTe_2/Fe_3GeTe_2$ van der Waals heterostructure. *Nat. Commun.* **11**, 3860 (2020).
42. Lenz, T. et al. Imaging topological spin structures using light-polarization and magnetic microscopy. *Phys. Rev. Appl.* **15**, 024040 (2021).
43. Tetienne, J.-P. et al. Quantum imaging of current flow in graphene. *Sci. Adv.* **3**, e1602429 (2017).
44. Ku, M. J. H. et al. Imaging viscous flow of the Dirac fluid in graphene. *Nature* **583**, 537–541 (2020).
45. Chen, H. et al. Revealing room temperature ferromagnetism in exfoliated $Fe_5GeTe_2$ flakes with quantum magnetic imaging. Preprint at https://arxiv.org/abs/2110.05314 (2021).

46. McLaughlin, N. J. et al. Quantum imaging of magnetic phase transitions and spin fluctuations in intrinsic magnetic topological nanoflakes. Preprint at https://arxiv.org/abs/2112.09863 (2021).
47. Andersen, T. I. et al. Electron-phonon instability in graphene revealed by global and local noise probes. *Science* **364**, 154–157 (2019).
48. Kolkowitz, S. et al. Probing Johnson noise and ballistic transport in normal metals with a single-spin qubit. *Science* **347**, 1129–1132 (2015).
49. McCullian, B. A. et al. Broadband multi-magnon relaxometry using a quantum spin sensor for high frequency ferromagnetic dynamics sensing. *Nat. Commun.* **11**, 5229 (2020).
50. Finco, A. et al. Imaging non-collinear antiferromagnetic textures via single spin relaxometry. *Nat. Commun.* **12**, 767 (2021).
51. Du, C. et al. Control and local measurement of the spin chemical potential in a magnetic insulator. *Science* **357**, 195–198 (2017).
52. Burch, K. S., Mandrus, D. & Park, J.-G. Magnetism in two-dimensional van der Waals materials. *Nature* **563**, 47–52 (2018).
53. Zhang, X.-Y. *et al.* ac Susceptometry of 2D van der Waals magnets enabled by the coherent control of quantum sensors. *PRX Quantum* **2**, 030352 (2021).
54. Fuchs, G. D., Dobrovitski, V. V., Toyli, D. M., Heremans, F. J. & Awschalom, D. D. Gigahertz dynamics of a strongly driven single quantum spin. *Science* **326**, 1520–1522 (2009).
55. Gottscholl, A. et al. Room temperature coherent control of spin defects in hexagonal boron nitride. *Sci. Adv.* **7**, eabf3630 (2021).
56. van der Sar, T., Casola, F., Walsworth, R. & Yacoby, A. Nanometre-scale probing of spin waves using single electron spins. *Nat. Commun.* **6**, 7886 (2015).
57. Thiel, L. et al. Probing magnetism in 2D materials at the nanoscale with single-spin microscopy. *Science* **364**, 973–976 (2019).
58. Broadway, D. A. et al. Imaging domain reversal in an ultrathin van der Waals ferromagnet. *Adv. Mater.* **32**, 2003314 (2020).
59. Li, Q. et al. Patterning-Induced ferromagnetism of $Fe_3GeTe_2$ van der Waals materials beyond room temperature. *Nano Lett.* **18**, 5974–5980 (2018).
60. Flebus, B. & Tserkovnyak, Y. Quantum-impurity relaxometry of magnetization dynamics. *Phys. Rev. Lett.* **121**, 187204 (2018).
61. Wang, H. et al. Quantum sensing of spin transport properties of an antiferromagnetic insulator. Preprint at https://arxiv.org/abs/2011.03905 (2020).
62. Wang, H. et al. Characteristics and temperature-field-thickness evolutions of magnetic domain structures in van der Waals magnet $Fe_3GeTe_2$ nanolayers. *Appl. Phys. Lett.* **116**, 192403 (2020).
63. Brown, P. J., Ziebeck, K. R. A., Déportes, J. & Givord, D. Magnetic correlation in itinerant magnetic materials above $T_C$ (invited). *J. Appl. Phys.* **55**, 1881–1886 (1984).
64. Pelliccione, M. et al. Scanned probe imaging of nanoscale magnetism at cryogenic temperatures with a single-spin quantum sensor. *Nat. Nanotechnol.* **11**, 700–705 (2016).
65. Song, T. et al. Direct visualization of magnetic domains and moiré magnetism in twisted 2D magnets. *Science* **374**, 1140–1144 (2021).

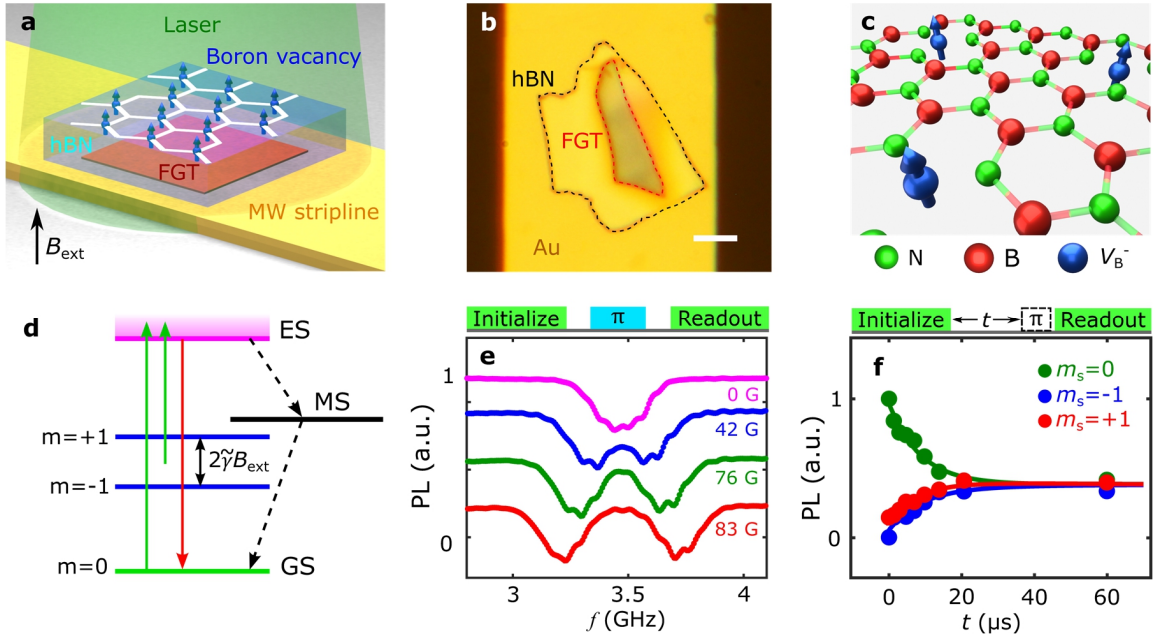

**Figure 1. Quantum sensing using $V_B^-$ spin defects in hexagonal boron nitride (hBN)**. **a** Schematic of a Fe$_3$GeTe$_2$(FGT)/hBN van der Waals heterostructure transferred onto an Au microwave stripline for wide-field magnetometry measurements. **b** Optical microscope image of a prepared FGT/hBN device. The FGT and hBN flakes are outlined with red and black dashed lines, respectively. The scale bar is 5 μm. **c** Schematic of $V_B^-$ spin defects (blue arrows) formed in a hexagonal crystalline structure with alternating boron (red) and nitrogen (green) atoms. A negatively charged boron atom vacancy $V_B^-$ is surrounded by three nitrogen atoms located in the nearest neighboring sites. **d** Energy level diagram of a $V_B^-$ spin defect and schematic illustration of optical excitation (green arrow), radiative recombination (red arrow), and nonradiative decay (black dotted arrow) processes between the ground state (GS), excited state (ES), and metastable state (MS). **e** Top panel: optical and microwave sequence of pulsed optically detected magnetic resonance (ODMR) measurements. Bottom panel: ODMR spectra of $V_B^-$ spin defects measured at a series of perpendicularly applied external magnetic fields $B_{ext}$. **f** Top panel: optical and microwave sequence of spin relaxometry measurements. Bottom panel: a set of spin relaxometry data of $V_B^-$ spin defects showing spin dependent photoluminescence measured as a function of delay time $t$. The external magnetic field is 185 G applied along the out-of-plane direction, and the measurement temperature is 295 K.

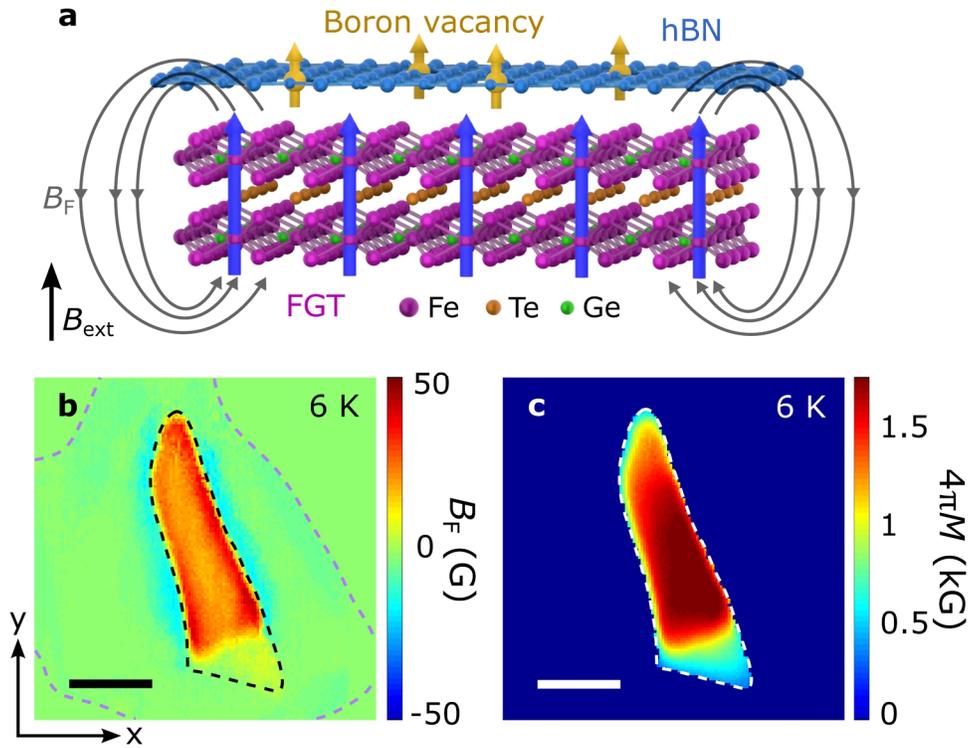

**Figure 2. Wide-field imaging of magnetization of an exfoliated FGT flake by $V_B^-$ spin defects in hBN. a** Schematic illustration of quantum sensing of local stray fields $B_F$ generated from FGT by proximate $V_B^-$ spin defects. **b, c** Two-dimensional maps of static stray field $B_F$ (**b**) and reconstructed magnetization $4\pi M$ (**c**) of an exfoliated FGT flake measured at 6 K with an external perpendicular magnetic field $B_{ext}$ of 142 G. The black and purple dashed lines in Fig. 2b outline the boundary of the FGT and hBN flake, respectively, and the scale bar is 5 μm.

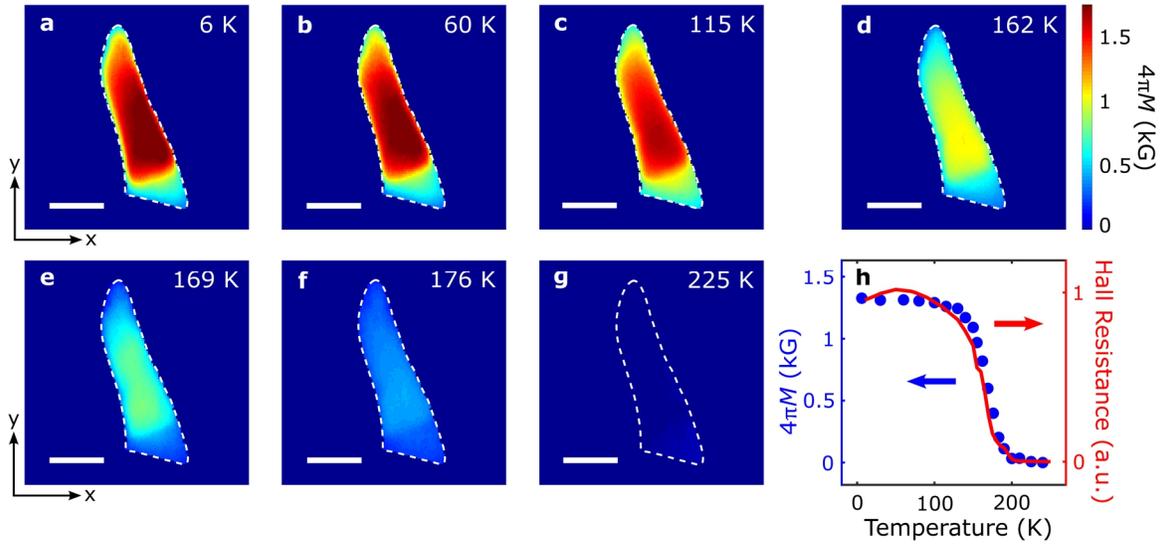

**Figure 3. Quantum imaging of temperature dependence of FGT magnetization. a-g** Reconstructed magnetization ($4\pi M$) maps of the FGT flake at $B_{ext}$ = 142 G and temperatures of 6 K (**a**), 60 K (**b**), 115 K (**c**), 162 K (**d**), 169 K (**e**), 176 K (**f**), and 225 K (**g**), respectively. The white dashed lines outline the boundary of the exfoliated FGT flake, and the scale bar is 5 µm. **h** Temperature dependence of spatially averaged magnetization of the FGT flake (blue points), in agreement with the variation behavior of the normalized Hall resistance presented in arbitrary units (a. u.) (red curve). The magneto-transport results were measured in a separate FGT flake with similar thickness.

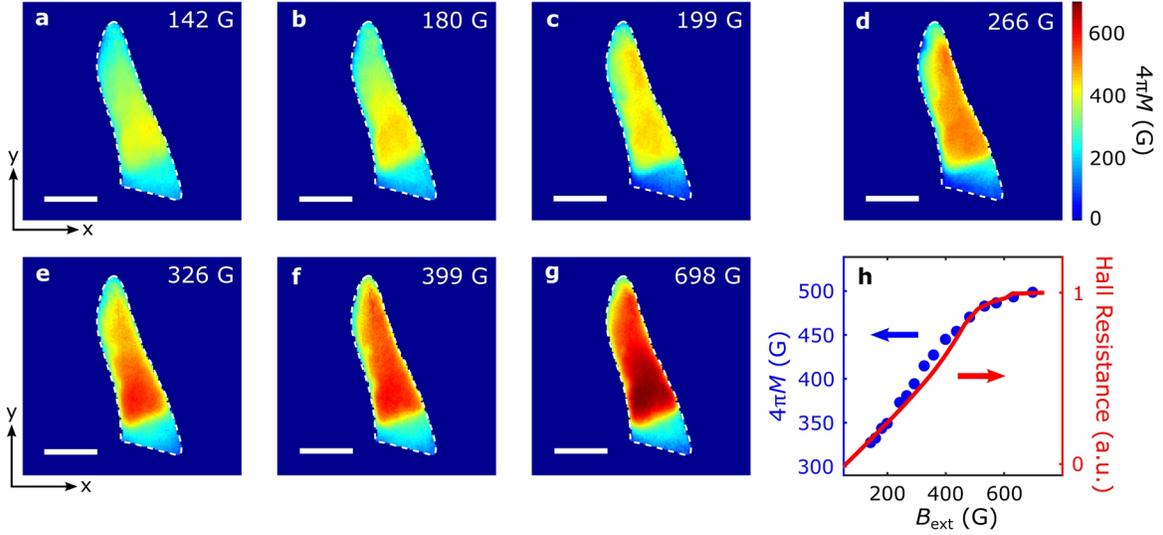

**Figure 4. Quantum imaging of field dependence of FGT magnetization. a-g** Reconstructed magnetization ($4\pi M$) maps of the FGT flake measured at 178 K with an external perpendicular magnetic field $B_{ext}$ of 142 G (**a**), 180 (**b**), 199 G (**c**), 266 G (**d**), 326 G (**e**), 399 G (**f**), and 698 G (**g**), respectively. The white dashed lines outline the boundary of the FGT flake, and the scale bar is 5 μm. **h** Field dependence of spatially averaged magnetization of the FGT flake (blue points), consistent with the variation behavior of the normalized Hall resistance presented in arbitrary units (a. u.) (red curve). The magneto-transport results were measured in a separate FGT flake with similar thickness.

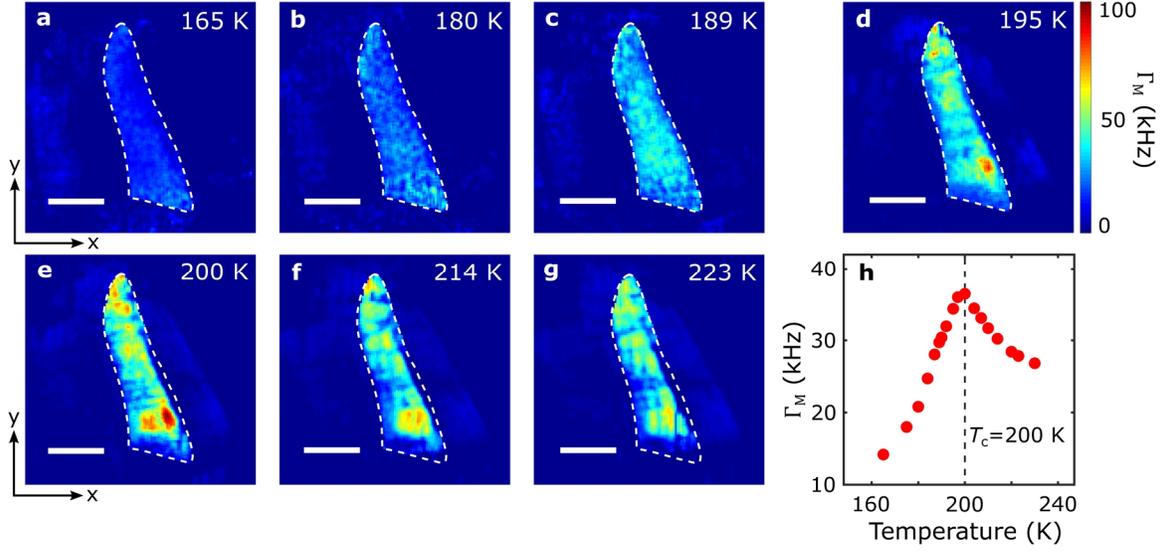

**Figure 5. Quantum imaging of spin fluctuations in an exfoliated FGT flake. a** Spin relaxation maps of $V_B^-$ spin defects measured at temperatures of 165 K (**a**), 180 K (**b**), 189 K (**c**), 195 K (**d**), 200 K (**e**), 214 K (**f**), and 223 K (**g**), respectively. The ESR frequency of $V_B^-$ spin defects $f_{ESR}$ is set to be approximately 1.9 GHz in these measurements with an external magnetic field $B_{ext}$ = 590 G. The white dashed lines outline the boundary of the FGT flake, and the scale bar is 5 μm. **h** Temperature dependence of the spatially averaged spin relaxation rate $\Gamma_M$ of $V_B^-$ spin defects located directly above the FGT flake. The black dashed line marks the Curie temperature of the FGT flake.